\documentclass[aps]{revtex4}

\usepackage{graphicx}
\usepackage{amsmath}
\usepackage{amssymb}

\newcommand{\ud}{\,{\mathrm d}}
\newcommand{\zc}{\overline{z}}
\newcommand{\wc}{\overline{w}}

\newcommand{\uiint}{\int\!\!\!\int}

\begin{document}
\title{A simple analytical description for the cross-tie domain wall structure.}
\author{Konstantin L. Metlov}
\affiliation{Institute of Physics ASCR, Na Slovance 2, Prague 8, 18221,
  Czech Republic} 
\email{metlov@fzu.cz}
\date{\today}
\begin{abstract}
  A closed form analytical expression for the magnetization vector
  distribution within the cross-tie domain wall in an isotropic
  ferromagnetic thin film is given. The expression minimizes the
  exchange energy functional exactly, and the magnetostatic energy by
  means of an adjustable parameter (wall width). The equilibrium value
  of the wall width and the film thickness corresponding to the
  transition between the N{\'e}el and the cross-tie walls are
  calculated. The results are compared with the recent experiments and
  are in good qualitative agreement.
\end{abstract}
\maketitle

The structure of magnetic domain walls (transition regions separating
magnetic domains) was in a focus of extensive research in 1930s -- 60s
and during the time a lot of useful results on the static structures
of various wall types as well as their dynamic properties were
obtained. Many of the results on the theory of the domain walls can be
found in books \cite{Hubert_book_walls,Aharoni_book}. In particular,
there are analytical expressions for the distribution of the
magnetization vector inside of the one-dimensional Bloch and N{\'e}el
domain walls, which are the starting point for calculating their
static and dynamic properties.

The cross-tie domain wall was also observed in a number of experiments
a that time (see Ref. \onlinecite{M61} and references therein) and
also with modern high resolution techniques
\cite{PCTZH93,SMZ96,LWBW96}, and are usual in thin and ultra-thin
magnetic films important for modern applications\cite{G01}. However,
there is no \cite{PCTZH93} (also see p. 163 in Ref.
\onlinecite{Aharoni_book}) closed form analytical expression for the
cross-tie wall structure.  This is, probably, due to the fact that
magnetization distribution even in straight cross-tie domain wall is
two-dimensional (unlike one-dimensional ones of Bloch and N{\'e}el
walls). It means, the structure of such a wall is defined by a system
of non-linear integral (due to the long-range dipolar interactions)
partial differential equations, and there is no way to reduce this
system to a single equation (as it was done for the N{\'e}el
wall\cite{RS71}).

Consider a thin film having the thickness $h$ made of soft (isotropic)
magnetic material, in the Cartesian coordinate system $X$, $Y$, $Z$
chosen in such a way that $0Z$ axis is perpendicular to the film
plane. The parameters of the material entering the calculation are the
exchange constant $C$, and the saturation magnetization constant
$M_S$.  If the film is thin enough, so that $h$ is of the order of a
few exchange lengths $L_E=\sqrt{C/M_S^2}$, the dependence of the
magnetization distribution on $Z$ can be neglected and the task
becomes essentially two-dimensional.  If we also neglect the dipolar
interaction for a while, the magnetization distribution
$\vec{M}(\vec{r})=M_S \, \vec{m}(\vec{r})$, $|\vec{m}|=1$,
$\vec{r}=\{X,Y\}$ is defined by the minimum of the exchange energy
functional:
\begin{equation}
  \label{eq:energy_exch}
  e^{\mathrm ex} =
  \frac{C}{2} h \uiint \sum_{i=x,y,z} (\vec{\nabla} m_i)^2 \ud^2 \vec{r}=
  C h \uiint \frac{4}{(1+w\overline{w})^2}
  \left(
    \frac{\partial w}{\partial z}
    \frac{\partial \overline{w}}{\partial\overline{z}}+
    \frac{\partial w}{\partial \overline{z}}
    \frac{\partial \overline{w}}{\partial z}
  \right) \ud^2 \vec{r},
\end{equation}
where the integration runs over all $X-Y$ plane. The last expression
is given using the parametrization of the magnetization vector field
by a complex function $w(z,\zc)$ of a complex variable $z=X + \imath
Y$, $\imath=\sqrt{-1}$, line over a variable means complex
conjugation, so that $m_x+\imath m_y = 2 w/(1+w\wc)$ and
$m_z=(1-w\wc)/(1+w\wc)$, $\partial/\partial z=(\partial/\partial X -
\imath \partial/\partial X)/2$, $\partial/\partial
\overline{z}=(\partial/\partial X + \imath \partial/\partial Y)/2$. It
is easy to see that the energy functional (\ref{eq:energy_exch}) is
scale-invariant, and the exchange energy of the magnetization
distribution $\vec{m}(\vec{r}/c)$ or $w(z/c,\zc /c)$, where $c$ is
real constant, is independent on $c$.  If we include the dipolar
interaction back into the picture this scale-invariance breaks. The
dipolar interaction will be considered in this paper to determine the
scale of the magnetization distribution obtained by minimizing
(\ref{eq:energy_exch}).

The system of Euler equations for the function $w(z,\zc)$ giving
extremum to the functional (\ref{eq:energy_exch}) is
\begin{equation}
  \label{eq:energy_euler}
  \frac{\partial}{\partial z}
  \left(\frac{\partial w}{\partial \overline{z}}\right) =
  \frac{2 \overline w}{1+w\overline{w}}     
  \frac{\partial w}{\partial z}
  \frac{\partial w}{\partial\overline{z}}.
\end{equation}

It is easy to see that any analytic (in a sense that the
Cauchy-Riemann conditions $\partial w/\partial \zc = 0$ are satisfied)
function of a complex variable satisfies the equation
(\ref{eq:energy_euler}).  The partial solution of this class was first
written as a rational polynomial in $z$ by Belavin and Polyakov
\cite{BP75} and can be thought as a superposition of vortices. There
are other solutions whose applications to the magnetism of small
cylindrical particles are given elsewhere\cite{M01_solitons}.

\begin{figure}[tbp]
  \begin{center}
    \includegraphics[scale=0.6]{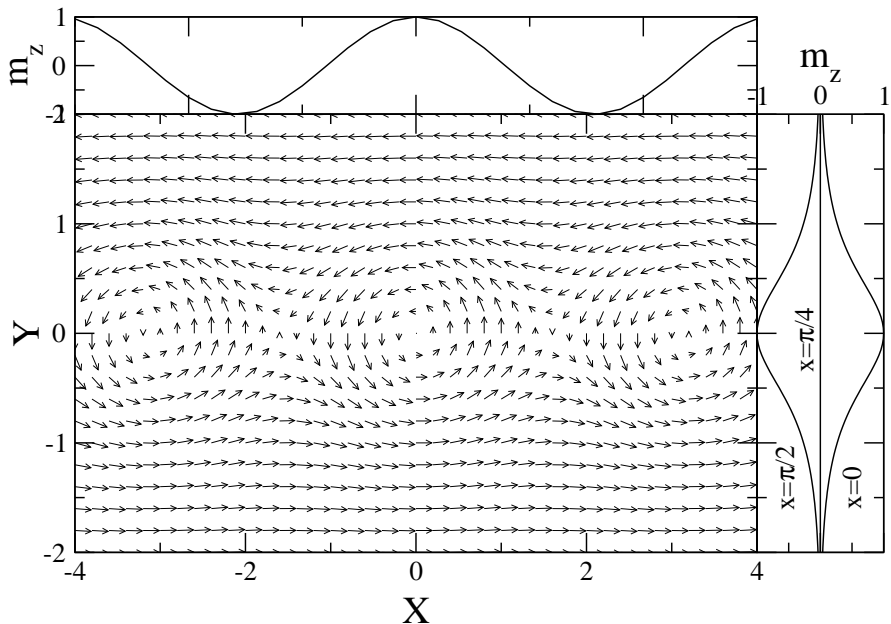}
    \caption{Distribution of the magnetization vector within the
      cross-tie domain wall having with $c=1$.}
    \label{fig:structure}
  \end{center}
\end{figure}

Consider the function 
\begin{equation}
  \label{eq:cross_tie_w}
  w_{C-T}(z)=i \tan(z/c) ,
\end{equation}
while the Cauchy-Riemann conditions are satisfied for this function,
and, therefore, it minimizes the functional (\ref{eq:energy_exch})
exactly, it is not a Belavin-Polyakov (BP) soliton because it can not
be represented as a rational polynomial in $z$ and, consequently, has
infinite energy when integrated over the whole space (while the energy
of BP solitons is always finite). Even though the total exchange
energy of the magnetization distribution $w_{C-T}$ is infinite, it is
extremal, because $w_{C-T}$ satisfies (\ref{eq:energy_euler}). Also,
the corresponding $\vec{m}=\{m_x,m_y,m_z\}= \{-\tanh(2 Y/c),\sin(2 X/
c)/\cosh(2 Y /c),\cos(2 X/ c)/\cosh(2 Y /c)\}$
(compare\cite{Landau_Lifshitz_wall}) plotted in
Fig.~\ref{fig:structure} has the topology of the magnetization
distribution in the cross-tie domain wall \cite{PCTZH93}.

The exchange energy per unit of the wall area is finite
\begin{equation}
  \label{eq:exchange_CT}
  \gamma^{\mathrm{ex}} = 
  \frac{e^{\mathrm{ex}}}{c \pi h} = 
  \frac{M_S^2 L_E^2}{c \pi}
  \int_{-c\pi/2}^{c\pi/2}\!\!\!\!\ud X
  \int_{-\infty}^{\infty}\!\!\!\!\ud Y \frac{4}{c^2 \cosh^2(2 Y/c)}
  = M_S^2 h \lambda^2
  \frac{4\pi}{\zeta},
\end{equation}
where dimensionless variables $\lambda=L_E/h$ and $\zeta=c \pi/h$ were
introduced. This energy term (\ref{eq:exchange_CT}) forces the wall to
expand infinitely as it has the smallest value $0$ only for
$c,\zeta\rightarrow\infty$.

The evaluation of the magnetostatic energy is straightforward, but a
little bit voluminous, it was done by calculating the density of
magnetic charges (having both volume and surface contributions),
solving the Poisson equation for the scalar magnetic potential in the
Fourier representation and performing the convolution of the density
of charges and the potential to obtain the energy. The final result
can be represented as
\begin{eqnarray}
  \label{eq:magnetostat_CT}
  \gamma^{\mathrm{dip}} & = & M_S^2 \, h f(\zeta), \\
  \label{eq:magnetostat_F}
  f(\zeta) & = & \frac{\zeta^2}{\pi} 
  \int_0^\infty \ud u 
  \int_0^\infty \ud v
  \frac{(u^2+v^2) \sin^2 (\pi v/\zeta)}{v^2 \cosh^2 (\pi u/2) (1+u^2+v^2)}.
\end{eqnarray}
One of integrals in $f(\zeta)$ can be taken analytically, the other is
rapidly converging and easy to take numerically. The representation
(\ref{eq:magnetostat_F}) was chosen for its compactness.  It is also
possible to build a simple analytical approximation for~$f(\zeta)$
\begin{eqnarray}
  \label{eq:magnetostatCT_appr}
  f(\zeta) & = & C_1 \zeta + C_2 (1 - e^{-2 \pi/\zeta}) \zeta^2 \\
  C_1 & = &\int_0^\infty \ud u \frac {\pi u}{(1+u^2)(1+\cosh \pi u)} = 0.17753\ldots\\
  C_2 & = &\int_0^\infty \ud u \frac{1}{4(1+u^2)^{3/2}\cosh^2 (\pi u/2)} = 0.12119\ldots\qquad ,
\end{eqnarray}
which is asymptotically correct for $\zeta \ll 1$ and produces the worst case
error at $\zeta\rightarrow\infty$ around $5\%$. 

\begin{figure}[tbp]
  \begin{center}
    \includegraphics[scale=0.6]{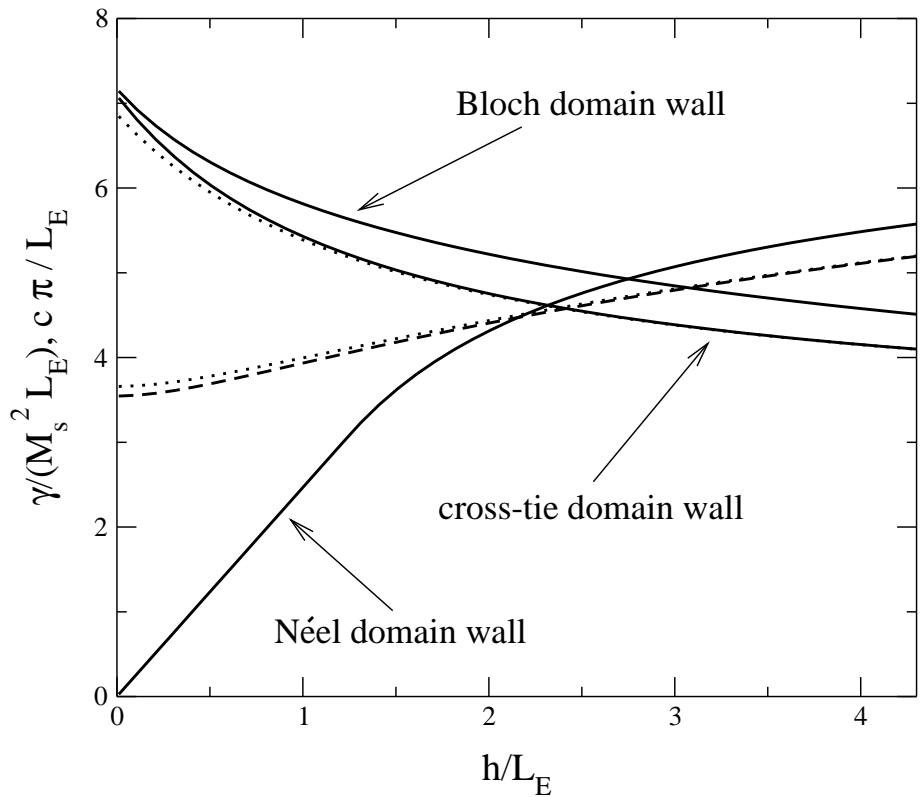}
    \caption{The normalized energy per unit wall area (solid lines)
      of the equilibrium cross-tie, N{\'e}el and Bloch domain walls as a
      function of film thickness. The dashed line shows the normalized width
      of the cross-tie wall. The dotted lines show the equilibrium cross-tie
      domain wall energy and width calculated using the approximate
      magnetostatic function.}
    \label{fig:energy-width}
  \end{center}
\end{figure}

The total energy per unit wall area
$\gamma=\gamma^{\mathrm{ex}}+\gamma^{\mathrm{dip}}$ was minimized
numerically with respect to the parameter $\zeta$, and the result is
shown in Fig.~\ref{fig:energy-width} in comparison to the energy per
unit area of the N{\'e}el and Bloch domain walls calculated according
to the model of Dietze and Thomas \cite{DT61}. As it could be expected
for films thicker than a certain value $h\approx 2.3 L_E$ the
cross-tie domain wall takes over the N{\'e}el one as the lowest energy
configuration.  Another interesting fact is that in the shown region
of film thicknesses (where roughly the assumption of uniformity of the
magnetization distribution along the thickness, $\partial
\vec{M}/\partial Z = 0$, can be expected to hold) the energy of the
cross-tie domain wall is lower than the energy of Bloch wall. This
suggests that the transition from the cross-tie wall to the Bloch one
takes place at larger thicknesses, where $\partial \vec{M}/\partial Z
= 0$ can not be safely assumed anymore.

As it is visible from Fig.~\ref{fig:energy-width} the approximate
expression for magnetostatic function (\ref{eq:magnetostatCT_appr})
after the energy minimization produces results almost
indistinguishable from the exact ones and having only a slight
deviations for extremely small film thicknesses. Thus, the expression
(\ref{eq:magnetostatCT_appr}) is safe to use instead of
(\ref{eq:magnetostat_F}) for most practical estimations of the
cross-tie wall magnetostatic energy.

The qualitative validity of the presented model is supported by the
good fit of the image signal (roughly proportional to the
magnetization vector component parallel to the wall plane) presented
in Fig. 10 of Ref.  \onlinecite{PCTZH93} to the empirically chosen
$\tanh(Y/\delta_w)$ function. In fact, $m_x(X,Y)=-\tanh(2 Y / c)$
exactly for the solution (\ref{eq:cross_tie_w}). Thus, the cross-tie
wall width, as defined in Ref. \onlinecite{PCTZH93}, is $\delta_w=
c/2$.  The permalloy film in that experiment had roughly $h/L_E=60nm
/18.2 nm \approx 3$.  Reading from Fig.~\ref{fig:energy-width} we get
the theoretical width of the cross-tie wall $\delta_w= 4.8 L_E/(2\pi)
\approx 14nm$, which is of the order of the measured
value\cite{PCTZH93} of $42 \pm 10 nm$.  It is expected that better
quantitative correspondence can be obtained for thinner films and also
by including the anisotropy energy (which is small for permalloy) into
the minimization.

The presented simple model for the magnetization distribution in the
cross-tie domain wall can be a starting point for analytical
estimations of its static and dynamic properties, especially for short
wavelength excitations on the cross-tie wall background, where the
fact that (\ref{eq:cross_tie_w}) minimizes the exchange energy
functional exactly can lead to significant simplifications. It can be
also useful for interpretation of MFM images, and for supplying
initial conditions for numerical finite-element micromagnetic
simulations.

This work was supported in part by the Grant Agency of the Czech
Republic under projects 202/99/P052 and 101/99/1662. I would like to
thank Vladimir Kambersk{\'y} and Vitaly Zablotskii for reading the
manuscript and many valuable discussions.


\begin{thebibliography}{10}
\bibitem{Landau_Lifshitz_wall}
The structure of the Landau-Lifshits wall is $\vec{m}=\{-\tanh(2 Y/c), 0,
  1/\cosh(2 Y /c)\}$.
\end{thebibliography}
\end{document}